\documentclass[9pt,twocolumn,twoside]{pnas-new}

\templatetype{pnasresearcharticle} 

\setboolean{displaywatermark}{false}

\title{Radiative heating achieves the ultimate regime of thermal convection}

\author[a,1]{Simon Lepot}
\author[a,b]{S\'ebastien Auma\^itre} 
\author[a,2]{Basile Gallet}

\affil[a]{Service de Physique de l'Etat Condens\'e, CEA Saclay, CNRS UMR 3680, Universit\'e Paris-Saclay, 91191 Gif-sur-Yvette, France.}
\affil[b]{Laboratoire de Physique, ENS de Lyon, UMR-CNRS 5672, 46 all\'ee d'Italie, 69007 Lyon, France}

\leadauthor{S. Lepot} 

\significancestatement{Turbulent convection is ubiquitous in geophysical and astrophysical contexts: it drives winds in the atmosphere and currents in the ocean, it generates magnetic fields inside planets and stars, and it triggers supernova explosions inside collapsing stellar cores. In many such natural flows, convection is driven by the absorption of incoming radiation (light or neutrinos). We designed an experiment to reproduce such radiatively driven convection in the laboratory. In contrast with convection driven by heating and cooling plates, our radiative heating setup achieves the ``ultimate'' regime of turbulent convection, which is the one relevant to many natural flows. Such experiments can yield the constitutive laws of turbulent convection, to be implemented into geophysical and astrophysical models.}

\authordeclaration{The authors declare no conflict of interest.}
\equalauthors{\textsuperscript{1} All the authors contributed equally to this work.}
\correspondingauthor{\textsuperscript{2} To whom correspondence should be addressed. E-mail: basile.gallet\@cea.fr}


\begin{abstract}
The absorption of light or radiation drives turbulent convection inside stars, supernovae, frozen lakes and the Earth's mantle. In these contexts, the goal of laboratory and numerical studies is to determine the relation between the internal temperature gradients and the heat flux transported by the turbulent flow. This is the constitutive-law of turbulent convection, to be input into large-scale models of such natural flows. However, in contrast with the radiative heating of natural flows, laboratory experiments have focused on convection driven by heating and cooling plates: the heat transport is then severely restricted by boundary layers near the plates, which prevents the realization of the mixing-length scaling-law used in evolution models of geophysical and astrophysical flows. There is therefore an important discrepancy between the scaling-laws measured in laboratory experiments and those used eg. in stellar evolution models. Here we provide experimental and numerical evidence that radiatively driven convection spontaneously achieves the mixing-length scaling regime, also known as the ``ultimate'' regime of thermal convection. This constitutes the first clear observation of this regime of turbulent convection. Our study therefore bridges the gap between models of natural flows and laboratory experiments. It opens an experimental avenue for a priori determinations of the constitutive laws to be implemented into models of geophysical and astrophysical flows, as opposed to empirical fits of these constitutive laws to the scarce observational data.
\end{abstract}

\dates{This manuscript was compiled on \today}
\doi{\url{www.pnas.org/cgi/doi/10.1073/pnas.XXXXXXXXXX}}

\newcommand{\cor}[1]{{#1}}

\begin{document}

\maketitle
\thispagestyle{firststyle}
\ifthenelse{\boolean{shortarticle}}{\ifthenelse{\boolean{singlecolumn}}{\abscontentformatted}{\abscontent}}{}

\dropcap{T}hermal convection drives natural flows in the atmosphere, in the oceans and in the interior of planets and stars. The resulting turbulence controls the convective heat transport, the typical wind speed in the atmosphere, the ability of planets and stars to produce magnetic field, and the triggering of supernova explosions inside collapsing stellar cores.
The cornerstone setup to study thermal convection is the Rayleigh-B\'enard one (RB), in which fluid is heated from below by a hot plate and cooled from above by a cold one. One then relates the convective heat transport enhancement to the temperature difference between the plates: 
one seeks a power-law relation $Nu \sim Ra^{\gamma}$, where the Nusselt number $Nu$ represents the dimensionless heat flux and the Rayleigh number $Ra$ characterizes the internal temperature gradients \cite{Malkus,Spiegel63,Kraichnan,Spiegel}. The goal is to determine the scaling-laws that govern the fully turbulent regime: these are the ``constitutive-laws'' of thermal convection, to be implemented eg. into simpler evolution models of astrophysical objects.

\begin{SCfigure*}[\sidecaptionrelwidth][t]
\centering
\includegraphics[width=13.5 cm]{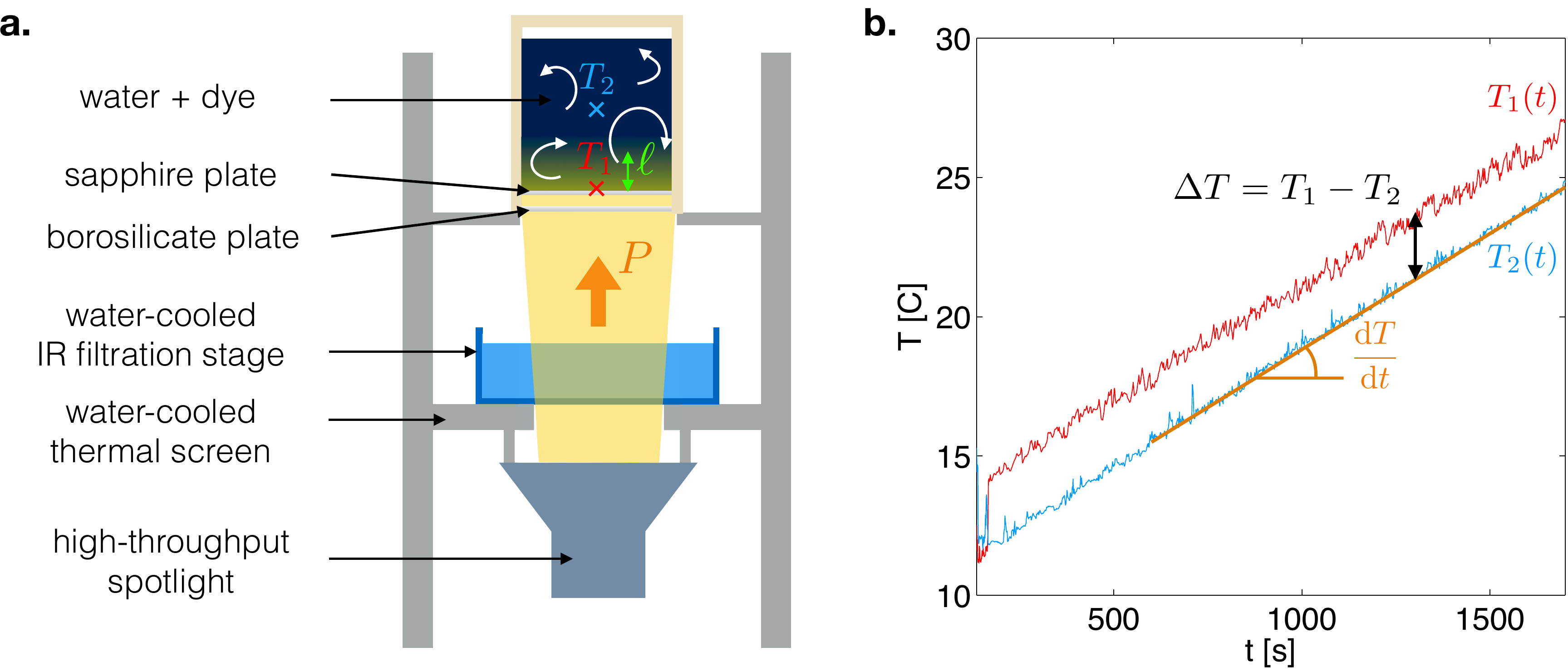} 
   \caption{\textbf{a. Radiatively driven convection in the laboratory.} A powerful spotlight shines at an experimental cell containing a mixture of water and dye. The light is absorbed over a height $\ell$ inversely proportional to the dye concentration. We measure the internal temperature gradients using two thermocouples: $T_1$ touches the bottom sapphire plate while $T_2$ is at at mid-depth. \newline {\bf b. Examples of time series.} The two temperature signals increase linearly with time, while a quasi-stationary temperature gradient $\Delta T= T_1-T_2$ is rapidly established. The slope $\mathrm{d}T/\mathrm{d}t$ gives direct access to the transported heat flux.}
\end{SCfigure*}

However, in spite of decades of investigations of the Rayleigh-B\'enard setup, the asymptotic regime of turbulent heat transport remains strongly debated and an outstanding challenge of nonlinear physics and turbulence research \cite{Chavanne97,Niemela,Chavanne01,Alhers,Roche,He}. \cor{Indeed, although the interior flow is strongly turbulent and transports heat very efficiently, this heat first has to be diffused across the boundary layers near the top and bottom plates. The temperature gradient is then confined to these lazy boundary layers \cite{Malkus}: standard dimensional analysis arguments give a heat transport exponent $\gamma=1/3$, while the measured values are typically $\gamma \simeq 0.3 \pm 0.03$.}
These measurements are in stark contrast with the prediction of a much more efficient mixing-length scaling regime of thermal convection -- the so-called ``ultimate'' regime -- where the heat flux is controlled by turbulence only \cite{Spiegel63,Kraichnan,Spiegel}. In this fully turbulent regime, viscosity and diffusivity become irrelevant. \cor{One can then introduce either a mixing length over which blobs of convecting fluid typically rise and fall, or directly ``turbulent'' values of viscosity and thermal diffusivity. Regardless of the details of the model, the key point is that these newly introduced quantities are independent of the molecular diffusion coefficients $\kappa$ and $\nu$ (but possibly depend on their ratio $Pr$, see below), and so is the heat flux: simple dimensional analysis then leads to $\gamma=0.5$. 
In the following we simply refer to the value $\gamma=0.5$ as the mixing-length or ``ultimate'' scaling-regime of thermal convection. 
The reader should keep in mind that it is the value of $\gamma$ that we investigate in the present study, more than the well-foundedness of the mixing-length assumptions, which are known to be questionable in many respects \cite{Spiegel}.} When extrapolated to the extreme parameter values of geophysical and astrophysical flows, \cor{the difference between the $\gamma = 0.3$ boundary-layer regime and the $\gamma = 0.5$ mixing-length regime translates into variations by orders of magnitude of the dimensionless heat flux or temperature gradients.}

There is a considerable controversy over the possible experimental detection of the mixing-length regime in the RB setup \cite{Chavanne97,Niemela,Chavanne01,Alhers,Roche,He}: most experiments report an increase in the exponent $\gamma$ at the highest achievable Rayleigh numbers, but a clear ``ultimate'' regime with $\gamma=0.5$ has yet to be observed in both experiments and direct numerical simulations (DNS). 
Insightful strategies have therefore been designed to minimize the role of the boundaries \cite{Shen,Gibert,Tisserand,Wei,Xie}. For instance, the use of rough plates disrupts the boundary layer structure and enhances the heat transport significantly, with $\gamma \simeq 0.5$ over some finite range of Rayleigh number. However, recent studies have shown that this effect is significant at moderate Rayleigh numbers only, while for strong imposed temperature gradients and fixed geometry the system settles back into the boundary-layer controlled regime \cite{Wettlaufer,Zhu,Rusaouen}.
\cor{Global rotation offers another way to alter the boundary-layer dynamics and increase the exponent $\gamma$ \cite{Julien96,Julien12}, albeit for an intermediate range of $Ra$ only, and with a heat flux that is smaller than its non-rotating counterpart \cite{King}.}

A clear observation of the mixing-lengh regime is all-the-more desirable in that it is the scaling-law currently used in astrophysical contexts. As an example, stellar evolution models are to be integrated over the lifespan of a star, which makes direct numerical simulations of the fluid dynamics prohibitively expensive and requires a parametrization of the convective effects \cite{Vitense,Spiegel,Miller,Shaviv}. Such parametrizations are based on the mixing-length scaling-law, and in the absence of experimental data to back it up one needs to fit the details of the convective model to the scarce observational data, which strongly restricts the predictive power of the whole approach. There is therefore an important research gap between laboratory experiments of turbulent convection on the one side, and parametrizations of geophysical and astrophysical convection on the other.

The present study bridges this gap by proposing an innovative experimental strategy leading to the mixing-length regime of thermal convection. It is based on the following observation: in contrast with Rayleigh-B\'enard convection, many natural flows are driven by a flux of light or radiation, instead of heating and cooling plates. A first example is the mixing of frozen lakes in the spring, due to solar heating of the near-surface water \cite{Farmer,Bengtsson,Jonas}. A second example is stellar interiors \cite{Featherstone} and the sun in particular, inside of which heat radiated by the core is transferred through the radiative zone before entering the convective one \cite{Spiegel92,Garaud,Christensen}. A related third example is convection in the Earth's mantle, which may be internally driven by radioactive decay \cite{Davaille,Limare}. Finally, the powering of supernova explosions by neutrino absorption in the collapsing stellar core provides an additional astrophysical example \cite{Herant,Janka,Radice}.

In this Letter, we report on an experimental setup showing that radiative heating spontaneously achieves the mixing-length regime of thermal convection. Indeed, radiative heating differs drastically from the RB setup, with important consequences for the transported heat flux: heat is input directly inside an absorption layer of finite extension $\ell$. When this absorption length is thicker than the boundary layers, radiative heating allows us to bypass the boundary layers and heat up the bulk turbulent flow directly.

\cor{The combination of radiation and convection is also ubiquitous in atmospheric physics. Solar radiation heats up the ground, which in turn emits black-body radiation in the infrared. This outgoing radiation is absorbed by CO$_2$ and water vapour, each layer of atmosphere absorbing part of the IR flux and emitting its own black-body radiation. Solving this problem leads to the radiative equilibrium solution for the atmospheric temperature profile. However, the lower part of this radiative profile is strongly unstable to convective motion. The effect of such convection is to restore the mean adiabatic lapse rate in the lower part of the atmosphere (the troposphere). This is the equivalent of saying that, in the bulk of turbulent Boussinesq convection, the temperature field is well-mixed and approximately independent of height. While this crude modeling of convection is well satisfied by the atmospheric data in the bulk of the troposphere, it fails in the atmospheric boundary layer (ABL). Models of the ABL indeed require parametrizations of the turbulent convective fluxes to reproduce the observed temperature profiles \cite{Garratt}. The simplest configuration to study this problem corresponds to a constant temperature surface below an atmosphere subject to volumic (infrared) cooling \cite{Tomkins}. Although this is the most generic situation in the ABL, occasionally the hot ground can induce infrared heating in the first hundreds of meters of atmosphere, in which case the fluid is subject to both volumic heating and cooling \cite{Deardorff,Kondratyev}. The experiment described further could provide a simple laboratory model for these two situations.}


Figure 1a provides a sketch of the experimental setup: a high-throughput projector shines at a cylindrical cell which has a transparent bottom plate. The cell contains a homogeneous mixture of water and dye. Dye absorbs the incoming light over a typical height $\ell$. Through Beer-Lambert law \cite{Bouguer}, this leads to a source of heat that decays exponentially away from the bottom boundary over a height $\ell$: the local heating rate is proportional to $\exp(-z/\ell)$, with $z$ the vertical coordinate measured upwards from the bottom plate. By changing the concentration of the dye, we can tune the thickness $\ell$ of the heating region. For large dye concentration, we heat up the fluid in the immediate vicinity of the bottom plate, in a similar fashion to the RB setup. By contrast, for low dye concentration we heat up the bulk turbulent flow directly, therefore bypassing the boundary layers. We can thus produce both RB-type heating and bulk radiative heating within the same experimental device. As described in the SI Appendix, the practical implementation of radiative heating requires extreme care: first, the choice of the dye is critical, as it must have a uniform absorbance over the visible spectrum. Second, the powerful spotlight has an efficiency of roughly $20 \%$, with most of the input electrical power being turned into heat. Special attention must be paid to avoiding parasitic heating of the fluid, through a combination of water-cooled thermal screens and IR filtration stages.

\begin{figure}[!ht]
    \centerline{\includegraphics[width=9 cm]{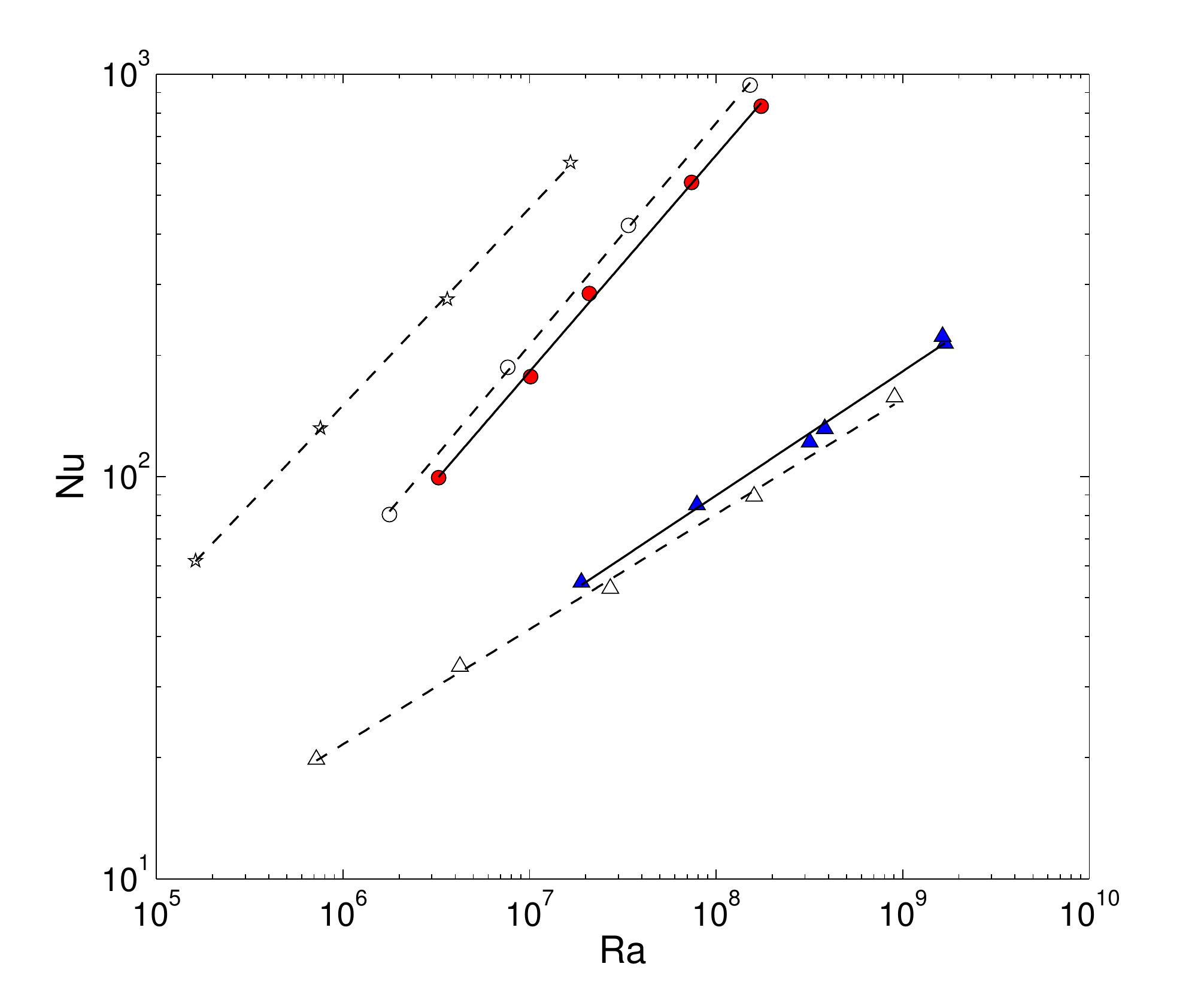} }
    \centerline{\includegraphics[width=9 cm]{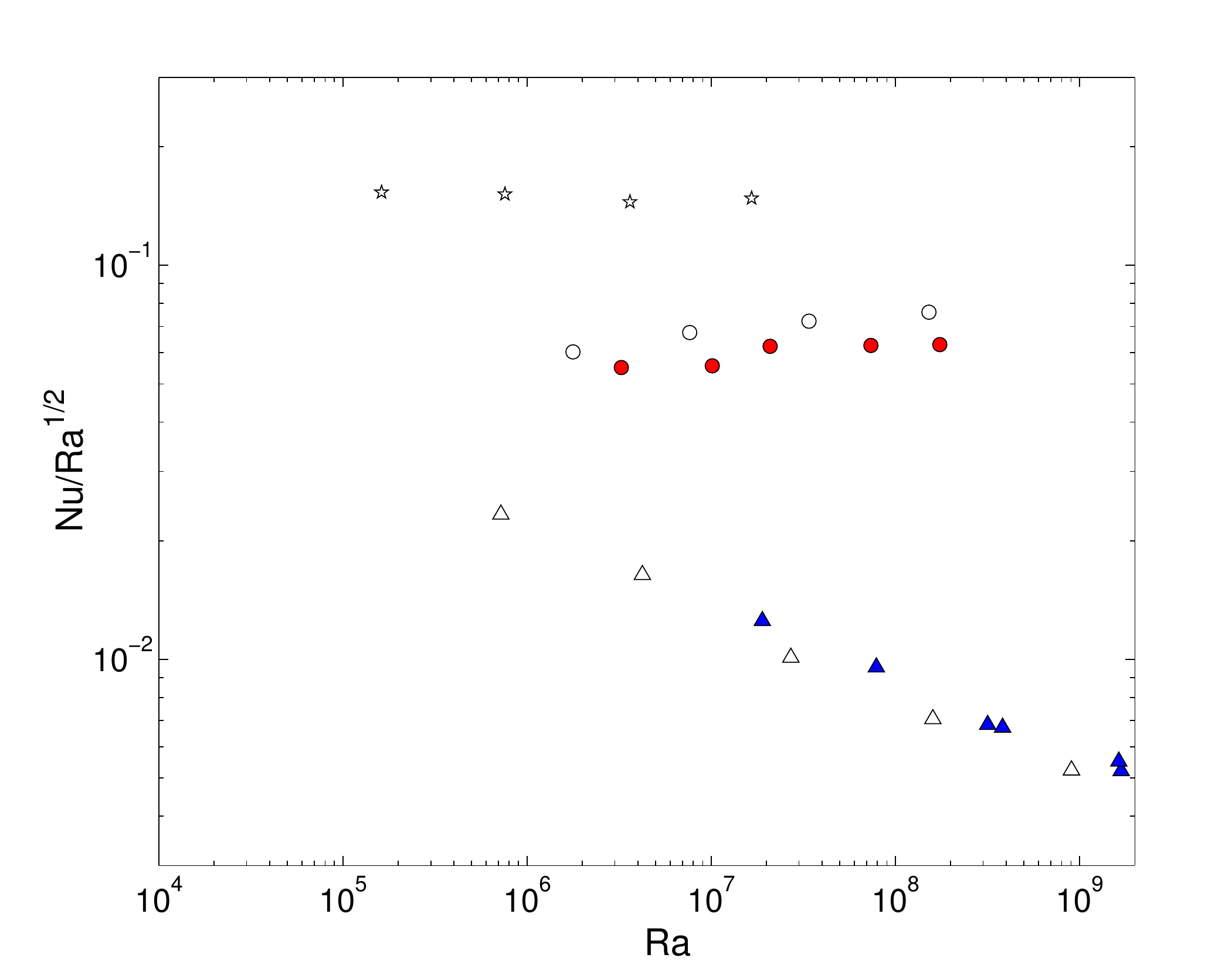} }
   \caption{\textbf{Nusselt number as a function of the Rayleigh number.} $\triangle$: Case I, corresponds to the RB limit. Filled blue triangles are experimental data with $\ell/H \leq 10^{-4}$, while empty triangles are DNS with a fixed flux bottom boundary condition. $\bullet$: Case II, $\ell/H=0.05$. Filled red circles are experimental data while empty circles are DNS. \cor{$\star$: Additional DNS performed with $\ell/H=0.1$ and $Pr=1$.} The lines are power-law fits to each curve. For case I, the fitted exponent is respectively $\gamma=0.31$ for experiments (solid line) and $\gamma=0.29$ for DNS (dashed line). For case II, the fitted exponent is $0.54$ for experiments (solid line) and $0.55$ for DNS (dashed line). \cor{The additional $Pr=1$ DNS give $\gamma=0.49$. The bottom panel shows the compensated Nusselt number: it varies by less than $20\%$ for the Case II data, and by less than 5\% for the additional $Pr=1$ DNS.}}
\end{figure}

Another key aspect of the experiment is to avoid boundary layers at the cooling side. Traditionally, studies of internally heated convection consider cooling at a solid isothermal boundary \cite{Kulacki}. The resulting boundary layers near the cold plate then control the heat transport efficiency and lead to scaling-laws similar to those of standard Rayleigh-B\'enard convection \cite{Goluskin}. The novel approach developed herein deviates from standard experiments: we run the experiment in quasi-stationary state, the fluid being radiatively heated without cooling mechanism. 
The resulting temperature field increases linearly with time. On top of this linear drift, the turbulent flow develops some stationary internal temperature gradients. At the mathematical level, this drifting situation is exactly equivalent to the fluid being radiatively heated and uniformly cooled at a rate equal and opposite to the heating power (see SI Appendix): the stationary internal temperature gradients measured experimentally are those of a fluid that is both internally heated and internally cooled, therefore bypassing both the heating and cooling boundary layers. This drift method is therefore a useful experimental tool to bypass the cooling boundary layers, but it also corresponds to practical situations: frozen lakes radiatively heating up in the spring and collapsing stellar cores heated by a flux of neutrinos. In a similar fashion, the secular cooling of planet interiors is routinely modelled as a uniform heating term in theoretical studies of planetary convection \cite{Gubbins,Aubert,Landeau}.

\section*{Results}

We characterize the temperature field using two precision thermocouples, one touching the bottom sapphire plate and the second one at mid-height, both being centered horizontally. Figure 1b presents the typical time series recorded during an experimental run. The common drift of the two sensors gives access to the heat flux $P$ transferred from the projector to the fluid: $P=\rho C  H $d$T$/d$t$, where $C$ is the specific heat capacity of water, $\rho$ the density, $H$ the fluid depth and d$T$/d$t$ denotes the slope of the drift. The latter is constant in the range 17 $^o$C to 27 $^o$C, which indicates that thermal losses are negligible. On top of this drift is a temperature difference $\Delta T$ between the two sensors, corresponding to internal temperature gradients. This temperature difference is stationary over the range 17 $^o$C to 27 $^o$C, which corresponds to typically 100 turnover times of the convective flow. We average $\Delta T$ over time to compute the Rayleigh and Nusselt numbers, defined as: 
\begin{equation}
Ra = \frac{\alpha g \left<\Delta T\right> H^3}{\kappa \nu} \, , \quad Nu=\frac{P  H}{\lambda \left<\Delta T \right>} \, ,
\end{equation}
where $\alpha$ denotes the thermal expansion coefficient of water, $g$ the acceleration of gravity, $\lambda$ the thermal conductivity, $\kappa$ the thermal diffusivity, $\nu$ the kinematic viscosity \cor{and $\left< \cdot \right>$ the time average}. \cor{An alternate choice could have been to define a Rayleigh number based on the heat flux, $Ra_P=\alpha g P H^4 / \nu \kappa \lambda$, which is a natural control parameter of the experiment. However, we prefer to stick to the standard $Ra$ and $Nu$ to make the comparison to RB studies more straightforward. The scaling laws discussed here are easily translated in terms of $Ra_P$ using the relation $Ra_P=Ra \times Nu$.}

Figure 2 reports the experimental curves $Nu$ versus $Ra$ for two limiting cases. Case I corresponds to $\ell=5 \mu$m, i.e., heating located in the immediate vicinity of the bottom plate, with $\ell/H\leq 10^{-4}$. Case II corresponds to $\ell/H=0.05$, with significant heating directly inside the bulk turbulent flow, away from the boundary layers. As expected, the curve $Nu$ vs $Ra$ obtained in case I is very similar to the Rayleigh-B\'enard situation. The data displays a clear power law, and a fit gives the exponent $\gamma=0.31$. This value is compatible with standard studies of Rayleigh-B\'enard convection and corresponds to a heat flux strongly restricted by the boundary layer near the bottom plate \cite{Niemela,Malkus}. By contrast, the data of case II shows significantly larger values of the Nusselt number. More importantly, the power-law exponent is now $\gamma=0.54$: radiative heating allows us to bypass the boundary layers and obtain a clear signature of the mixing-length -- or ``ultimate'' -- scaling regime. 

\begin{SCfigure*}[\sidecaptionrelwidth][t]
\centering
\includegraphics[width=14 cm]{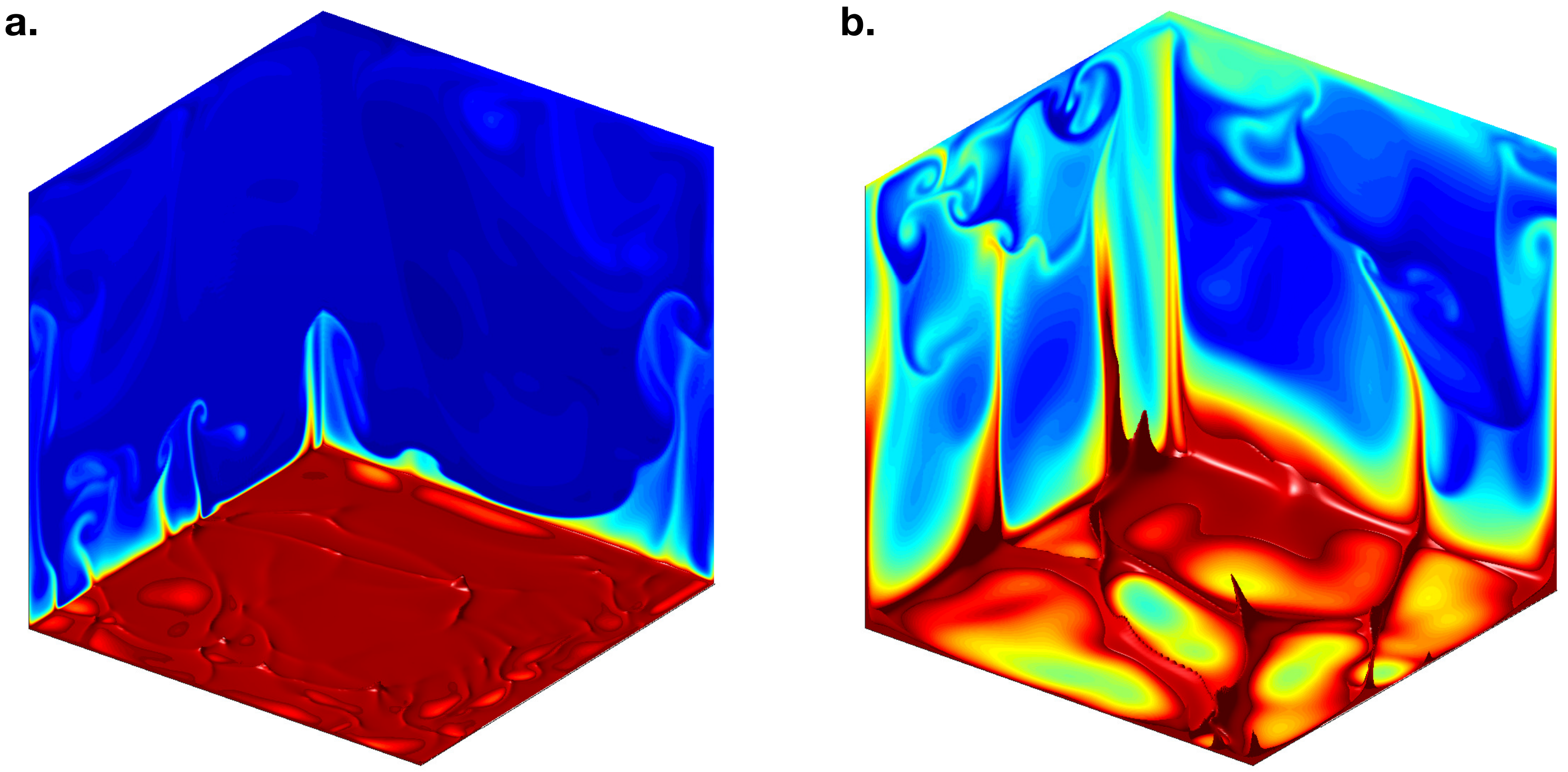} 
   \caption{\textbf{Temperature field from DNS.} We represent the temperature field at the boundaries, together with the isosurface of half-maximum temperature, for case-I and case-II DNS performed at the same heating power. The colorscale ranges from blue for the minimum temperature inside the domain to red for the maximum one.  \textbf{a.} Case I: the temperature gradients are located very near the bottom boundary, with narrow plumes seldom penetrating the bulk of the fluid domain. \textbf{b.} Case II: the region of warm fluid extends more in the vertical direction, with taller and wider plumes penetrating the bulk turbulent region.}
\end{SCfigure*}

To confirm the experimental data, we have performed Direct Numerical Simulations of radiatively driven convection. The domain is a 3D cube with stress-free side walls and no-slip boundary conditions at the top and bottom boundaries. We first fix the Prandtl number to the value of water at 20 $^o$C, $Pr=7$. To simulate case I above, the sidewalls and top boundaries are thermally insulating, while we impose a fixed flux boundary condition at the bottom. This corresponds to radiative heating in the limit $\ell/H \to 0$. To simulate case II, all the boundaries are thermally insulating, and the fluid is internally heated through a source term that decays exponentially with height. In both cases, cooling is ensured by a uniform sink term modeling the drift in the experimental situation (see the SI Appendix for a proof of this exact equivalence). We extract the Rayleigh and Nusselt numbers once the simulations reach a statistically steady state. The corresponding data points are shown in figure 2: the agreement with the experimental data is excellent. Indeed, the numerical values of the exponent $\gamma$ are 0.29 in case I and 0.55 in case II, confirming the experimental values with a precision of a few percent. 
\cor{The slight departure of the case II exponent from $0.5$ may originate from a small finite-Rayleigh-number viscous correction. To test this hypothesis, we have performed additional DNS using $Pr=1$ and $\ell/H=0.1$. The corresponding values of the Nusselt number are shown in figure 2, and a power-law fit leads to the exponent $\gamma=0.49$, even closer to the $0.5$ theoretical value. To further illustrate this point, in the bottom panel of figure 2 we show the compensated Nusselt number $Nu/Ra^{1/2}$ as a function of the Rayleigh number. As expected, for the RB case this compensated Nusselt number decreases rapidly with increasing Rayleigh number. By contrast, for the case-II data this quantity is almost constant, with less than 20\% variations over two decades in $Ra$. The additional DNS performed with $Pr=1$ exhibit even smaller variations, the compensated Nusselt number being constant within 5\% over two decades in Rayleigh number.}

\section*{Discussion}

\cor{Another open question concerning the ultimate regime of thermal convection is the behaviour of the Nusselt number with respect to the Prandtl number. For RB convection, Kraichnan predicted that $Nu$ should scale as $Pr^{1/2}$ for small Prandtl number, and as $Pr^{-1/4}$ for ``moderate'' Prandtl numbers. The $Pr^{1/2}$ scaling is the most fashionable one among astrophysicists. It is the one obtained when one asks for the heat flux to be independent of viscosity first, and then of thermal diffusivity \cite{Spiegel}. However, depending on how one takes the double limit $(\nu,\kappa) \to (0,0)$, it may very well be that the heat flux retains some dependence on the ratio $Pr=\nu/\kappa$: see the moderate Prandtl number scaling of Kraichnan \cite{Kraichnan}, or reference  \cite{Gallet} for an analytical example of a flow that retains a dependence on the ratio of two dissipative coefficients as they go to zero. Some insight may also be obtained from rigorous upper bound theory: for RB convection, the best upper bound on the Nusselt number  scales as $Ra^{1/2}$ and is independent of the Prandtl number \cite{Howard,Doering}. The fact that these bounds cannot capture any $Pr^{1/2}$ dependence may be an indication that the Nusselt number is in fact independent of $Pr$, at least over some range of $Pr$. Coming back to the radiatively-heated setup, we ran a few additional DNS at $Pr=1$, $Pr=7$ and $Pr=20$, with constant $\ell/H$: the corresponding points fall onto the same $Nu \sim Ra^{1/2}$ curve, indicating that the Nusselt number seems to be independent of $Pr$ in this intermediate range of $Pr$, which includes some of the natural flows mentioned at the outset. The extensive description of the parameter space goes beyond the scope of the present Letter, and we defer it to a future publication.}


Figure 3 presents snapshots of the temperature field in the high-Rayleigh-number DNS. For case I, most of the temperature gradients are contained in a very thin boundary layer. These temperature gradients are strong, because only diffusion acts to evacuate the heat away from the heating region, where there is little or no advection. The high temperature values only seldom penetrate the bulk of the domain through the emission of narrow plumes of warm fluid. By contrast, in case II the high temperature regions extend significantly further from the bottom plate, and penetrate the bulk turbulent flow through taller and wider thermal plumes. \cor{The horizontally averaged temperature profiles for these two snapshots are provided in the SI Appendix, Fig. S1: for case I, the temperature is homogeneous in the bulk, with a sharp boundary layer near the bottom wall. For case II, the temperature decreases less rapidly with height, but we can still identify a small thermal boundary-layer near the bottom wall. Defining its thickness $\delta_{th}$ as the height at which the mean temperature profile drops by $\left<\Delta T\right>/2$, the DNS data gives $\delta_{th}/H \simeq 0.015 \ll \ell/H$.  This inequality indicates that heat is input mostly into the bulk turbulent flow, which results in $\Delta T$ being weaker} than in case I: the Nusselt number is much larger and the overall efficiency of the heat transport is greatly enhanced. 

We conclude by stressing once again that the current theory of stellar evolution strongly relies on the mixing-length scaling regime of turbulent convection \cite{Vitense,Barker,Miller}: the simplest implementations of mixing-length theory consist in fitting the mixing length to the scarce observational data, which limits the predictive power of the whole approach. Instead, one would like to determine the constitutive laws of turbulent convection {\it a priori}, based on experimental and numerical studies in idealized geometries. A pre-requisite is that such experiments should achieve the mixing-length regime of thermal convection. While the mixing-length regime has yet to be clearly observed in the standard Rayleigh-B\'enard system, our radiative experiment provides the first clear observation of this scaling regime. The present study therefore puts the mixing-length scaling theory on a firmer footing by reconciling it with laboratory experiments. It paves the way for the {\it a priori} experimental determination of the convective parametrizations to be input into stellar evolution models, potentially including non-locality and/or overshooting at an internal boundary \cite{Miller,Shaviv}.

\showmatmethods{} 

\acknow{This research is supported by the European Research Council (ERC) under grant agreement FLAVE 757239, and by PALM ``Excellence laboratory'' ANR-10-LABX-0039. The authors thank J. Guilet, T. Foglizzo, C.R. Doering, D. Goluskin and V. Bouillaut for insightful discussions.}

\showacknow{} 

\bibliography{pnas-sample}

\begin{thebibliography}{99}


\bibitem{Malkus} W.V.R. Malkus, The heat transport and spectrum of thermal turbulence, {\it Proc. R. Soc. Lond. A}, {\bf 225}, 196-212 (1954).

\bibitem{Spiegel63} E.A. Spiegel, A generalization of the mixing-length theory of thermal convection, {\it ApJ} {\bf 138}, 216 (1963).

\bibitem{Kraichnan} R.H. Kraichnan, Turbulent thermal convection at arbitrary Prandtl number, {\it Phys. Fluids} {\bf 5}, (1962).

\bibitem{Spiegel} E.A. Spiegel, Convection in stars I. Basic Boussinesq convection, {\it Annu. Rev. Astron. Astrophys.}, {\bf 9}, 323-352 (1971).

\bibitem{Chavanne97} X. Chavanne et al., Observation of the ultimate regime in Rayleigh-B\'enard convection, {\it Phys. Rev. Lett.} {\bf 79}, (1997).

\bibitem{Niemela} J.J. Niemela, L. Skrbek, K.R. Sreenivasan, R.J. Donnelly, Turbulent convection at very high Rayleigh numbers, {\it Nature} {\bf 404}, 837-840 (2000).

\bibitem{Chavanne01} X. Chavanne et al., Turbulent Rayleigh-B\'enard convection in gaseous and liquid He, {\it Phys. Fluids} {\bf 13}, (2001).

\bibitem{Alhers} G. Alhers, S. Grossmann, D. Lohse, Heat transfer and large-scale dynamics in turbulent Rayleigh-B\'enard convection, {\it Rev. Mod. Phys.} {\bf 81}, (2009).

\bibitem{Roche} P.-E. Roche et al., On the triggering of the ultimate regime of convection, {\it New. J. Phys.} {\bf 12}, (2010).

\bibitem{He} X. He et al., Transition to the ultimate state of turbulent Rayleigh-B\'enard convection, {\it Phys. Rev. Lett.} {\bf 108}, (2012).




\bibitem{Shen} Y. Shen, P. Tong, K.-Q. Xia, Turbulent convection over rough surfaces, {\it Phys. Rev. Lett.} {\bf 76}, 908 (1996).

\bibitem{Gibert} M. Gibert et al., High-Rayleigh-Number convection in a vertical channel, {\it Phys. Rev. Lett.} {\bf 96}, 084501 (2006).

\bibitem{Tisserand} J.-C. Tisserand et al., Comparison between rough and smooth plates within the same Rayleigh-B\'enard cell, {\it Phys. Fluids} {\bf 23}, (2011).

\bibitem{Wei} P. Wei et al., Heat transport properties of plates with smooth and rough surfaces in turbulent thermal convection, {\it J. Fluid Mech.} {\bf 740}, 28-46 (2014).

\bibitem{Xie} Y.-C. Xie, K.-Q. Xia, Turbulent convection over rough plates with varying roughness geometries, {\it J. Fluid Mech.} {\bf 825}, 573-599 (2017).

\bibitem{Wettlaufer} S. Toppaladoddi, S. Succi, J.S. Wettlaufer, Roughness as a route to the ultimate regime of thermal convection, {\it Phys. Rev. Lett.} {\bf 118}, 074503 (2017).

\bibitem{Zhu} X. Zhu et al., Roughness-facilitated local 1/2 scaling does not imply the onset of the ultimate regime of thermal convection, {\it Phys. Rev. Lett.} {\bf 119}, 154501 (2017).

\bibitem{Rusaouen} E. Rusaou\"en, O. Liot, B. Castaing, J. Salort, F. Chill\'a, Thermal transfer in Rayleigh-B\'enard cell with smooth or rough boundaries, {\it J. Fluid Mech.} {\bf 837}, 443-460 (2018).

\bibitem{Julien96} K. Julien, S. Legg, J. McWilliams, J. Werne, Rapidly rotating turbulent Rayleigh-B\'enard convection, {\it J. Fluid Mech.} {\bf 322}, 243-273 (1996).

\bibitem{Julien12} K. Julien, E. Knobloch, A.M. Rubio, G.M. Vasil, Heat transport in low-Rossby-number Rayleigh-B\'enard convection, {\it Phys. Rev. Lett.} {\bf 109}, 254503 (2012).

\bibitem{King} E. King, S. Stellmach, J. Noir, U. Hansen, J. Aurnou, Boundary layer control of rotating convection systems, {\it Nature} {\bf 457} (2009).




\bibitem{Vitense} E. Vitense, Die wasserstoffkonvecktionszone der sonne, {\it Z. Astrophysik}, {\bf 32}, 135-164 (1953).



\bibitem{Miller} M.M. Miller Bertolami et al., On the relevance of bubbles and potential flows for stellar convection, {\it MNRAS}, {\bf 457}, 4441-4453 (2016).

\bibitem{Shaviv} G. Shaviv, E.E. Salpeter, Convective overshooting in stellar interior models, {\it ApJ}, {\bf 184}, 191-200 (1973).




\bibitem{Farmer} D. Farmer, Penetrative convection in the absence of mixing, {\it Quart. J. R. Met. Soc.} {\bf 101}, 869-891 (1975).

\bibitem{Bengtsson} L. Bengtsson, Mixing in ice-covered lakes, {\it Hydrobiologia} {\bf 322}, 91-97 (1996).

\bibitem{Jonas} T. Jonas, A. Terzhevik, D. Mironov, A. W\"uest, Radiatively driven convection in an ice-covered lake investigated using temperature microstructure technique, {\it J. Geophys. Res.} {\bf 108}, (2003).

\bibitem{Featherstone} N.A. Featherstone, B.W.Hindman, The spectral amplitude of stellar convection and its scaling in the high-Rayleigh-number regime, {\it ApJ} {\bf 818}, 1 (2016).



\bibitem{Spiegel92} E.A. Spiegel, J.-P. Zahn, The solar tachocline, {\it Astron. Astrophys.} {\bf 265}, 106-114 (1992).

\bibitem{Garaud} P. Garaud, Dynamics of the solar tachocline - I. An incompressible study, {\it Mon. Not. R. Astron. Soc.} {\bf 329}, 1-17 (2002).


\bibitem{Christensen} J. Christensen-Dalsgaard, The current state of solar modeling, {\it Science} {\bf 272}, 5266 (1996).




\bibitem{Davaille} A. Davaille, F. Girard, M. Le Bars, How to anchor hotspots in a convecting mantle? {\it Earth Planet. Sc. Lett.} {\bf 203}, 621-634 (2002).

\bibitem{Limare} A. Limare et al., Microwave-heating laboratory experiments for planetary mantle convection {\it J. Fluid Mech.} {\bf 777}, 50-67 (2015).


\bibitem{Herant} M. Herant, W. Benz, S. Colgate, Postcollapse hydrodynamics of SN 1987A: two-dimensional simulations of the early evolution, {\it ApJ} {\bf 395}, (1992).

\bibitem{Janka} H.-T. Janka, E. M\"uller, Neutrino heating, convection, and the mechanism of type-II supernova explosion, {\it Astron. Astrophys.} {\bf 306}, (1996).

\bibitem{Radice} D. Radice et al., Neutrino-driven convection in core-collapse supernovae: high-resolution simulations, {\it ApJ} {\bf 820}, (2016).


\bibitem{Garratt} J.R. Garratt, Review: the atmospheric boundary layer, {\it Earth. Sci. Rev.} {\bf 37}, 89-134 (1994).

\bibitem{Tomkins} A.M. Tomkins, G.C. Craig, Radiative-convective equilibrium in a three-dimensional cloud-ensemble model, {\it Q. J. R. Meteorol. Soc.} {\bf 124}, 2073-2097 (1998).

\bibitem{Deardorff} J.W. Deardoff, Three-dimensional numerical study of the height and mean structure of a heated planetary boundary layer, {\it Boundary Layer Meteoreology} {\bf 7}, 1, 81-106 (1974).

\bibitem{Kondratyev} K.Y. Kondratyev, On the October 1970 - March 1972 complex atmospheric energetics experiment (CAENEX) results, {\it Abridged report to the joint organizing committee of GARP}, (1972).

\bibitem{Bouguer} P. Bouguer, Essai d'optique sur la gradation de la lumi\`ere, {\it Claude Jombert}, 16-22 (1729).


\bibitem{Kulacki} F.A. Kulacki, R.J. Goldstein, Thermal convection in a horizontal fluid layer with uniform volumetric energy sources, {\it J. Fluid Mech.}, {\bf 55}, 271-287 (1972).

\bibitem{Goluskin} D. Goluskin, Internally heated convection and Rayleigh-B\'enard convection, {\it Springer}, (2016).





\bibitem{Gubbins} D. Gubbins, D. Alfe, G. Masters, G.D. Price, M.J. Millan, Can the Earth's dynamo run on heat alone?, {\it Geophys. J. Int.} {\bf 155}, 2 (2003).


\bibitem{Aubert} J. Aubert, S. Labrosse, C. Poitou, Modelling the Palaeo-evolution of the geodynamo, {\it Geophys. J. Int.} {\bf 179}, 3 (2009).


\bibitem{Landeau} M. Landeau, J. Aubert, Equatorially asymmetric convection inducing a hemispherical magnetic field in rotating spheres and implications for the past Martian dynamo, {\it Phys. Earth Planet. Inter.} {\bf 185}, (2011).



\bibitem{Gallet} B. Gallet, W.R. Young, A two-dimensional vortex condensate at high Reynolds number, {\it J. Fluid Mech.} {\bf 715}, 359-388 (2013).


\bibitem{Howard} L.N. Howard, Heat transport by turbulent convection, {\it J. Fluid Mech.} {\bf 17}, 405-432 (1963).


\bibitem{Doering} C.R. Doering, P. Constantin, Variational bounds on energy dissipation in incompressible flows. III. Convection, {\it Phys. Rev. E} {\bf 53}, 5957 (1996).



\bibitem{Barker} A.J. Barker, A.M. Dempsey, Y. Lithwick, Theory and simulations of rotating convection, {\it ApJ}, {\bf 791} (2014).

\end{thebibliography}

\end{document}